\documentclass[conference]{IEEEtran}
\IEEEoverridecommandlockouts
\usepackage{cite}
\usepackage{amsmath,amssymb,amsfonts}
\usepackage{graphicx}
\usepackage{textcomp}
\usepackage{xcolor}
\usepackage[hidelinks]{hyperref}

\def\BibTeX{{\rm B\kern-.05em{\sc i\kern-.025em b}\kern-.08em
    T\kern-.1667em\lower.7ex\hbox{E}\kern-.125emX}}

\begin{document}

\title{Machine Learning for Designing `Undesignable' Metal-Organic Frameworks}

\author{
\IEEEauthorblockN{Satya Kokonda}
\IEEEauthorblockA{Charter School of Wilmington\\
Wilmington, Delaware, USA\\
\href{https://orcid.org/0009-0004-0477-5847}{orcid.org/0009-0004-0477-5847}}
}

\maketitle

\begin{abstract}
Many crucial processes are too complex for computational modeling, requiring experimentation to identify promising materials. Here, a general methodology for application-specific material design is presented, with photocatalysis presented as a specific case study. Metal-Organic Frameworks (MOFs) are a subset of highly promising porous nanomaterials used in a variety of `unmodellable' applications. Reinforcement learning generated 60,000 novel MOFs optimized for CO/H\textsubscript{2}O selectivity. A predictor funnel system was created, iteratively removing low-scoring MOFs down to 10,986 potential candidates, improving computational efficiency by 276\%. Trained Crystal Graph Convolutional Neural Network (CGCNN) models predicted features for a fitness function incorporating stability, catalytic ability, material cost, sustainability, and adsorption, while allowing the inclusion of application-specific design criteria. This designed function provides a computational method to model photocatalytic performance, and filtered down to two promising MOFs which each pass a myriad of synthesis criteria: first, a Cr-based MOF with a photocatalyst score 230\% higher than the control; second, a Zn-based MOF that outperforms the best control material across all relevant metrics, demonstrating robustness against variable fitness functions. This work designed 20 materials, each 125\% better than the control for this application. Furthermore, analysis revealed insightful design patterns, such as the significant influence of metal cluster N262 on catalytic performance, providing a method for future work to narrow the chemical space. By incorporating industrially applicable features such as cost or stability of the material, this work successfully designs industrially promising materials for otherwise unmodellable processes such as drug delivery, while presenting a method for multi-objective optimization incorporating 260\% more features than prior work.
\end{abstract}


\section{Introduction}

\subsection{General Information}

Metal--organic frameworks (MOFs) have been identified as one of the ``Top Ten Emerging Technologies in Chemistry'' by the International Union of Pure and Applied Chemistry (IUPAC)~\cite{gomollon2019ten}. MOFs are porous nanomaterials characterized by modular structures composed of metal clusters and organic linkers, organized according to specific topologies~\cite{park2024inverse}. Their high surface area and structural tunability have made them promising candidates for diverse applications, including energy storage~\cite{chu2019metal}, drug delivery~\cite{gupta2019peg}, and catalysis~\cite{muller2008loading}.

Traditional computational approaches to MOF screening rely on time-intensive, force-field--based methods such as Universal Force Field (UFF)--driven computational theory~\cite{nazarian2015benchmarking,park2024inverse}. However, recent advancements in machine learning (ML) have significantly improved the scalability and efficiency of MOF property prediction, enabling rapid evaluation across large-scale material libraries. ML-driven methods have been applied to assess MOFs in applications such as adsorption-based heat pumps and chemical sensors~\cite{okur2020towards}.

Among these applications, photocatalytic MOFs are of growing interest due to their potential to facilitate solar-driven carbon dioxide (CO\textsubscript{2}) reduction. These materials convert CO\textsubscript{2} into value-added products such as carbon monoxide (CO), which is utilized in biomedical treatments for conditions including sepsis, sickle cell disease, and organ transplant complications~\cite{gao2023applications}. The photocatalytic process typically involves three sequential steps: (1) surface binding of CO\textsubscript{2} on the MOF surface, (2) cleavage of C--O bonds and formation of C--H bonds via electron and proton transfer, and (3) release of the final product~\cite{zhang2022high}.

\subsection{Problem Background}

In prior computational modeling studies, the majority of efforts have focused on optimizing a single material property using simulations such as Density Functional Theory (DFT)~\cite{nazarian2015benchmarking}, or, in more advanced cases, two properties using Pareto front analysis~\cite{deng2024multiscale}. While effective in certain domains, these approaches are inadequate for photocatalysis, which presents distinct challenges.

Photocatalytic systems involve complex, multi-step reaction mechanisms and dynamic surface interactions that are difficult to capture with current simulation techniques. Consequently, computational predictions often lack fidelity, and experimental validation remains both time-consuming and resource-intensive. Additionally, many machine learning or simulation-based design frameworks tend to optimize for a narrow definition of performance, frequently overlooking additional critical criteria~\cite{nazarian2015benchmarking}. This is especially limiting in photocatalytic applications, where overall material efficacy depends on a broad and interdependent set of structural, electronic, and thermodynamic features~\cite{park2024inverse}. As such, optimization strategies that rely solely on one or two design objectives risk producing materials that perform poorly in practical settings.

\subsection{Design Criteria}

We hypothesize that if machine learning was used to sift through a large corpus of MOF building blocks and instead model the application itself, maximizing a variety of features relevant to that application, it would be able to design a material for an otherwise unmodellable and ``undesignable'' process.

To evaluate the proposed workflow, it is applied to the design of a novel photocatalytic metal--organic framework (MOF) for CO\textsubscript{2} reduction, with the goal of outperforming existing materials in this domain. One of the primary concerns in computational screening is feature overfitting~\cite{redkar2023carbnn}, where the optimization algorithm assigns excessive weight to a narrow subset of material features, potentially ignoring other critical performance factors. To mitigate this, the workflow incorporates mechanisms to ensure a balanced evaluation across multiple properties.

This study leverages a MOF building block database that is over 104 times larger than the most comprehensive prior datasets~\cite{park2024inverse}, significantly expanding the design space. Given this scale, computational efficiency is a core priority to maintain feasibility during large-scale screening.

In addition to catalytic performance, the design process incorporates economic and sustainability criteria. Target materials are selected to minimize cost and environmental impact, prioritizing precursors that can be derived from waste sources such as electronic sludge~\cite{boukayouht2023sustainable}. Prior work has demonstrated the feasibility of recovering metal ions from battery and e-waste for use in MOF synthesis, and this study is the first to explicitly optimize for waste-derived components in the design process.

The proposed MOFs are also subject to constraints ensuring operational robustness, including thermal stability at temperatures up to 148\textdegree~C (300\textdegree~F)~\cite{power2010101}, resistance to structural degradation~\cite{fischer2018tuning}, and strong CO\textsubscript{2} adsorption under humid conditions~\cite{park2024inverse}.

Finally, the designed structures are evaluated for synthetic feasibility using established criteria to ensure their realizability in laboratory conditions~\cite{park2024inverse,fu2023mofdiff}.

\section{Methods}

\subsection{MOF Generation}

\begin{figure}[!t]
\centering
\includegraphics[width=\columnwidth]{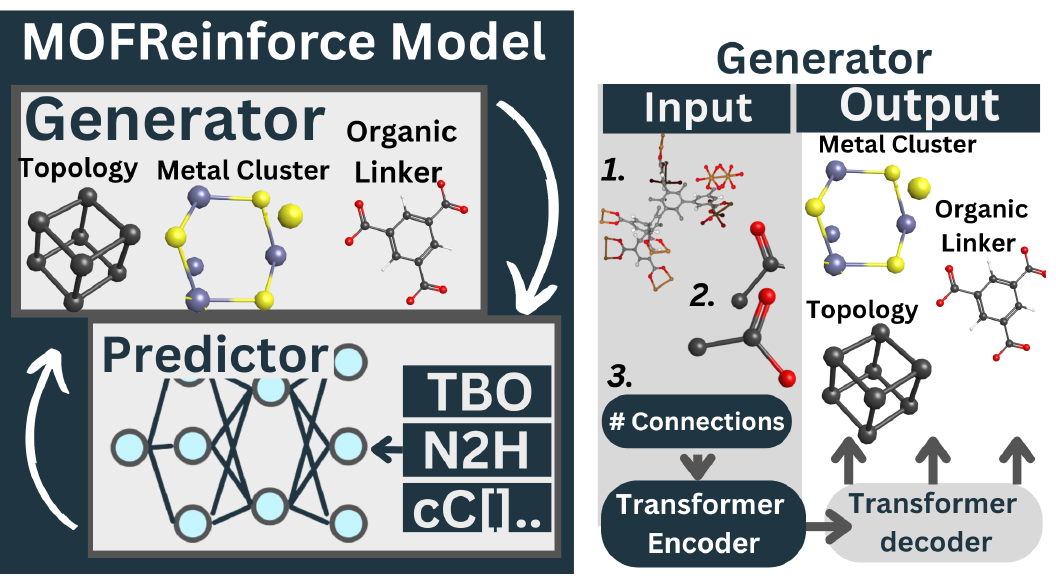}
\caption{Diagram of the MOFReinforce encoder--decoder architecture. The model is based on a Transformer framework and takes as input the metal cluster (MC), material scaffold, and the number of MC connections. It generates full MOF structure by predicting the topology, metal cluster, and organic linker. The generated MOFs are tokenized and passed through a neural network to estimate CO\textsubscript{2}/H\textsubscript{2}O selectivity. These predictions are then used as feedback to optimize the generator via reinforcement learning, guiding the synthesis of higher-performing materials.}
\label{fig:1}
\end{figure}

As shown in Fig.~\ref{fig:1}, a total of 60,000 metal-organic frameworks (MOFs) were generated using a modified version of the MOFReinforce model~\cite{park2024inverse}. For this application, the original model was adapted to produce six times its standard output volume. This enhancement was achieved through three key modifications to the model architecture and training process, enabling the generation of a significantly larger and more diverse dataset suitable for downstream machine learning tasks.

\begin{itemize}
\item \textbf{Stochastic Manipulation:} The model incorporates both biased and unbiased generators, each with a 50\% probability of selecting a value for the 128 slots that define the tokenized representation of a material. As a result, repeated inputs can yield significantly different material structures due to this controlled randomness.
\item \textbf{Noise Injection:} Minor variations were introduced into the organic linker scaffolds. Since these scaffolds are reused multiple times within a single structure, small changes can lead to amplified differences, effectively generating distinct materials from a common base.
\item \textbf{Set Reconfiguration:} In the original MOFReinforce framework, the output is determined by the size of the test set. By reallocating a portion of the validation set to the test set during training, the model's output capacity was expanded without compromising the training-validation integrity.
\end{itemize}

The selectivity of CO\textsubscript{2}/H\textsubscript{2}O is critical for the high water environment of power plants~\cite{barla2022process}.

\begin{figure}[!t]
\centering
\includegraphics[width=\columnwidth]{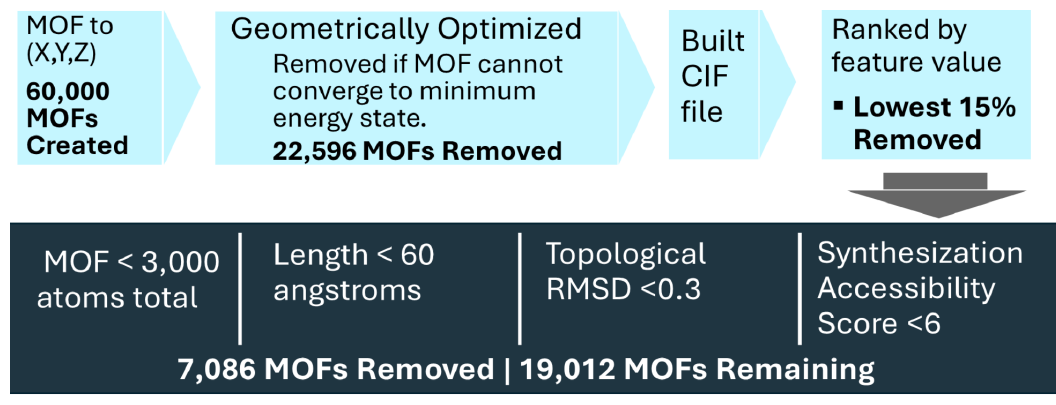}
\caption{Post-processing workflow applied following MOF generation. This includes structural validation, energy minimization, and elimination of low-performance MOFs. These steps ensure that only physically feasible and high-performing MOFs are retained for further analysis.}
\label{fig:2}
\end{figure}

As illustrated in Fig.~\ref{fig:2}, the initial post-processing phase involved optimizing the geometry of each molecular structure. This was done using the Universal Force Field (UFF)~\cite{rappe1992uff} method. UFF is widely used to ensure that molecular structures are in a stable and realistic configuration before further analysis, particularly in machine learning pipelines.

The optimized molecular structures were converted into the Crystallographic Information File (CIF) format using RDKit~\cite{rdkit2025}, an open-source toolkit for molecular data processing. CIF is a widely adopted standard for storing molecular and crystal structure information, particularly in materials science and chemistry. In this workflow (available at~\cite{satyak2024github}), various input formats---including JSON, tokenized representations, and $(x,y,z)$ coordinate data---were consistently translated into CIF. This standardized format ensures compatibility with graph neural networks and supports flexible integration of diverse data sources throughout the study.

To improve data quality and reduce computational load, the bottom 15\% of entries (ranked by selectivity) were removed. This filtering left 33,054 structures. The materials had to pass preliminary synthesis feasibility criteria, which also served to avoid excessive computational costs by removing candidates that were either too large (over 3,000 atoms) or too spatially complex (over 60~\AA\ in size)~\cite{park2024inverse}. After these steps, 19,012 metal-organic frameworks (MOFs) remained and were carried forward for further modeling and analysis.

\subsection{CGCNN Modification}

The Crystal Graph Convolutional Neural Network (CGCNN) represents crystalline materials as graphs, with atoms as nodes and interatomic bonds as edges. Convolutional layers are applied to these graphs to extract both local and global structural features, enabling the prediction of material properties directly from raw crystal data~\cite{xie2018crystal}. In this work, the original CGCNN model was extended to incorporate real-time performance tracking features, including Root Mean Squared Error (RMSE) and precision/recall metrics, enabling dynamic visualization of training progression and identification of optimal epoch counts via loss curve analysis.

Given that CGCNN was initially optimized for inorganic compounds such as oxides, sulfides, and intermetallics, architectural modifications were made to account for potential overfitting when applied to alternate crystalline systems. These adjustments resulted in an 18\% reduction in mean absolute error (MAE) and a 36\% improvement in computational efficiency. The fine-tuned model was subsequently employed for accurate material property prediction for the remainder of this work.

\subsection{Funnel System}

\begin{figure}[!t]
\centering
\includegraphics[width=\columnwidth]{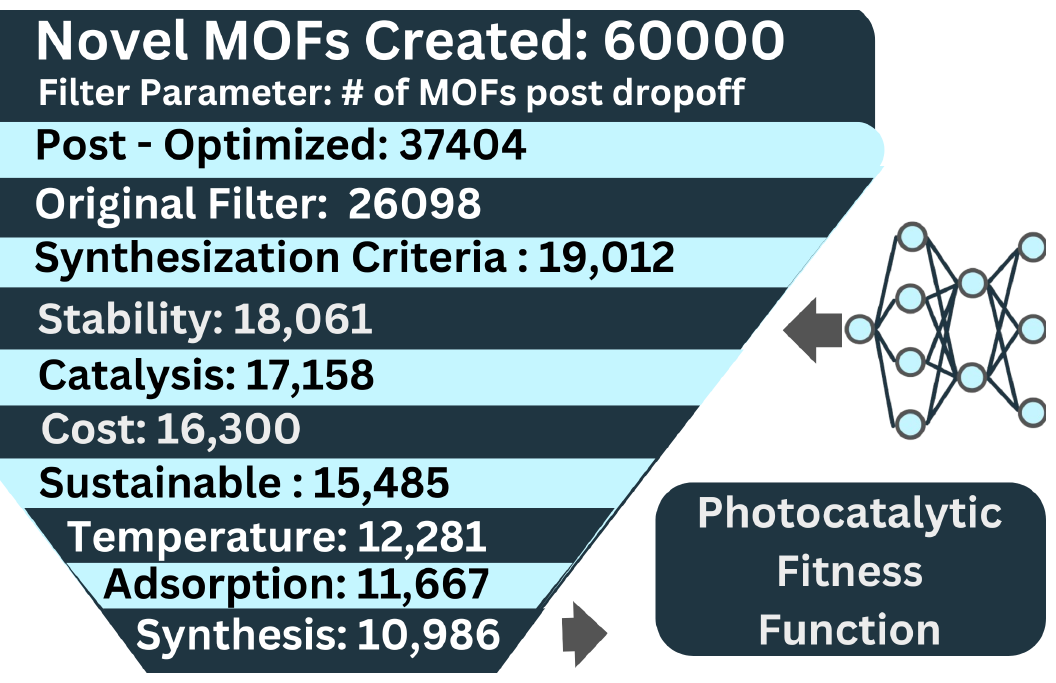}
\caption{Created chain-linked CGCNN funnel system for model prediction. Using 13 CGCNN modules, the compound properties were predicted, with the lowest 5\% of the MOFs being removed in the later layers. This reduces feature overfitting and improves computational efficiency by 276\%.}
\label{fig:3}
\end{figure}

An iterative funnel system of trained Crystal Graph Convolutional Neural Networks (CGCNNs)~\cite{xie2018crystal} was created to predict various features detailing photocatalytic success, as illustrated in Fig.~\ref{fig:3}. The funnel system, characterized by significant drop-off, balances all structural features, prevents overfitting, and reduces computational costs by eliminating MOFs demonstrating extremely low viability in any relevant feature.

Industrial viability of materials requires a high degree of structural stability, which is evaluated through two primary aspects. Water stability is critical, as flue gas from coal-fired power plants typically contains 20--23\% H\textsubscript{2}O, which can degrade unstable materials over time~\cite{barla2022process}. Additionally, mechanical stability---quantified by a high bulk modulus---ensures the material can withstand high-pressure gas environments while maintaining consistent catalytic performance across multiple operational cycles~\cite{fischer2018tuning}. A dataset for MOF water stability was generated via the MOFSimplify 2-class water stability program~\cite{nandy2022mofsimplify}, run on MOFReinforce-generated MOFs to have the input data closely match the expected inputs, improving prediction accuracy. Both the best Bulk-Modulus model pretrained from the CGCNN and a trained CGCNN on the aforementioned water-stability dataset were run on the remaining 19,012 MOFs. The lowest 5\% summed stability values were removed.

Given the strong similarities in the chemical process between photocatalytic CO\textsubscript{2} reduction and electrocatalytic CO\textsubscript{2} reduction, features shown to impact electrocatalytic success are extrapolated to photocatalytic success. Three pre-trained models (for Faradaic Efficiency, Free Energy, and Voltage Potential), along with the fitness function (1) for electrocatalytic CO\textsubscript{2} reduction, were utilized from CarbNN~\cite{redkar2023carbnn}:
\begin{equation}
\frac{\text{FaradaicEfficiency}}{5} - 5\cdot\text{FreeEnergy} - \frac{|\text{VP}|}{2}
\label{eq:carbnn}
\end{equation}
where VP denotes Voltage Potential.

Band gap expects an ideal range of 1.8--2~eV for photocatalytic CO\textsubscript{2} reduction for ideal light absorption~\cite{wu2017co2reduction}, compared to a `maximize is best' rule used for most other features. A function (2), a modified bell-curve output, was added to the electrocatalytic fitness function defined in CarbNN (Eq.~\ref{eq:carbnn}), with the bottom 5\% of MOFs trimmed off, resulting in 17,158 values remaining:
\begin{equation}
B(x) = 4.536 \cdot e^{-\frac{(x-1.9)^2}{150}}
\label{eq:bandgap}
\end{equation}

This work incorporates both material cost and sustainably sourced materials. Each atom's cost was found by creating a dataset, available at~\cite{satyak2024github}, for the MOFs found in the QMOF database, estimating the cost of each element in its common industrially used form. The sustainability dataset was created by providing a score for every atom that can be derived from waste materials~\cite{boukayouht2023sustainable} from the Quantum MOF (QMOF)~\cite{rosen2021machine} database, using an underrepresenting-the-majority-class system given that most MOFs demonstrated a sustainability score of 0 in the training set. The highest 5\% (to ensure lowered costs) MOFs by cost and the lowest 5\% MOFs by sustainability were then removed sequentially, leaving 15,485 materials remaining.

The temperature dataset was created from the MOFSimplify dataset~\cite{nandy2022mofsimplify}, with data augmentation from MOFSimplify-program-calculated MOFs to improve data quality. The criterion of stability above 148\textdegree~C (300\textdegree~F)~\cite{power2010101} left 12,281 compounds remaining.

Three key metrics are used to evaluate adsorption effectiveness, each aligned with a distinct phase of the target chemical reaction. Available Surface Area (ASA) reflects the capacity for simultaneous catalytic interactions. CO\textsubscript{2}/H\textsubscript{2}O selectivity prevents surface binding with non-target molecules (H\textsubscript{2}O). Lastly, Heat of Adsorption (between $-20$ and $-40$~kJ/mol) indicates optimal removal of the product after the process, supporting continuous reaction cycles~\cite{park2024inverse}. A database used for training the CGCNN to predict CO\textsubscript{2} Heat of Adsorption (HOA) and CO\textsubscript{2}/H\textsubscript{2}O selectivity was created from the MOFReinforce predictor information converted to CIF files. Existing literature does not specify an ideal heat of adsorption value, but based on prior work, an ideal range between $-20$ and $-40$~kJ/mol was approximated, with a slow taper in Eq.~\ref{eq:hoa} to incorporate viable materials found with HOA near $-60$~kJ/mol:
\begin{equation}
\begin{aligned}
g(x) &= e^{-\frac{(x+30)^2}{150}} \\
H(x) &= \frac{1}{f(-20)}\cdot \min\big(f(-20),\, f(x)\big)
\end{aligned}
\label{eq:hoa}
\end{equation}

Available Surface Area (ASA) was then trained from the MOSAEC~\cite{gibaldi2025mosaec} database, with the three normalized values added for the adsorption aspect, and the bottom 5\% were removed. A CGCNN was trained on QMOF~\cite{rosen2021machine} database synthesized MOFs, using a classification model. With a 0.5 cutoff, only 10,986 MOFs remained and were passed into the fitness function (Eq.~\ref{eq:fitness}).

\subsection{Modelling Photocatalytic Viability}

A key contribution of this study is the development of a fitness function---adapted from concepts in genetic algorithms~\cite{gustafson2019intelligent}---to quantitatively evaluate the photocatalytic performance of metal-organic frameworks (MOFs). This function, defined in Eq.~\ref{eq:fitness}, integrates both versatility and prior domain insights~\cite{redkar2023carbnn}, leveraging features generated from the funnel pipeline to reduce computational waste.

Multiplicative design in Eq.~\ref{eq:fitness} discourages feature overfitting by detrimentally reducing fitness in the presence of underperforming features, in contrast to previously utilized additive models (Eq.~\ref{eq:carbnn}). To benchmark the system, seven experimentally verified MOF photocatalysts from existing literature were curated as a control group. These were evaluated using the same CGCNN module and the proposed fitness metric to validate comparative performance and robustness.

\begin{equation}
\begin{aligned}
&\Big[(\text{BulkMod} + \text{WaterStability}) \cdot \frac{\sqrt{\text{Sustainability}+1}}{\sqrt{\text{Cost}+1}}\Big] \\
&\cdot \Big[\text{BandGap} + \frac{\text{Fara}}{5} - \text{FreeEnergy}\cdot 5 - \frac{|\text{VP}|}{2}\Big] \\
&\cdot \big[\text{ASA} + \text{CO}_2\text{HeatOfAdsorption} + \text{CO}_2\text{Selectivity}\big]
\end{aligned}
\label{eq:fitness}
\end{equation}
where Fara is Faradaic Efficiency, VP is Voltage Potential, and ASA is Available Surface Area.

\subsection{Data Analysis}

X-ray diffraction (XRD) is a standard technique for determining the crystallographic structure of materials. In this study, XRD patterns were simulated using the XRD module of the Python Materials Genomics (pymatgen) library~\cite{ong2013python}, with code available at~\cite{satyak2024github}. The analysis was applied to both newly designed MOFs generated in this work and reference synthesized structures sourced from the QMOF database~\cite{rosen2021machine}.

\section{Results}

\begin{figure}[!t]
\centering
\includegraphics[width=\columnwidth]{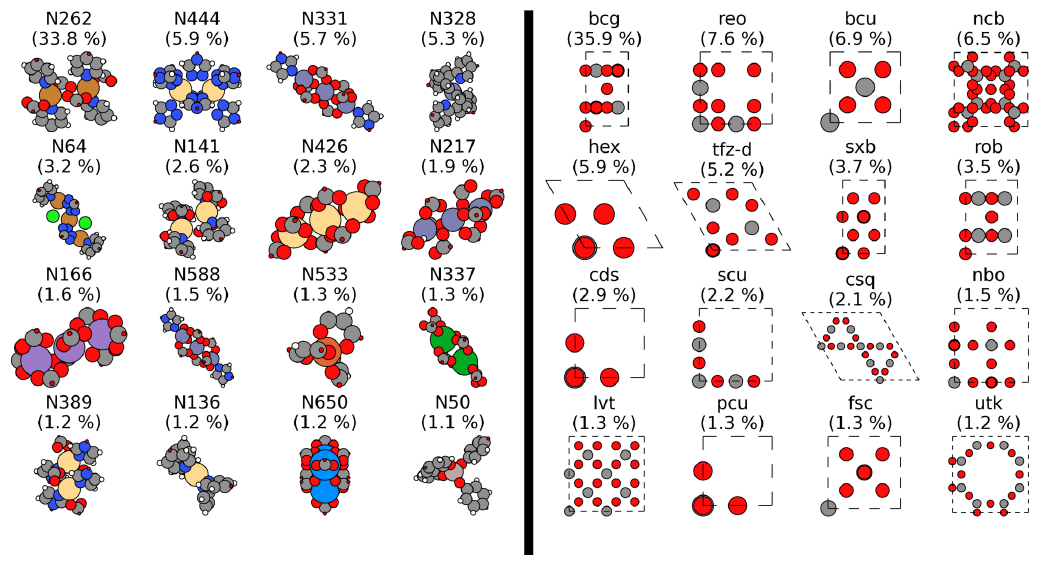}
\caption{The distribution of metal clusters and organic linkers among the top 1,000 MOFs, ranked by the fitness function in Eq.~\ref{eq:fitness}. In the topology diagrams, red nodes represent organic linker connections, while gray nodes denote metal clusters.}
\label{fig:4}
\end{figure}

Key structural features and design elements of high-performing photocatalytic materials were identified through analysis of a dataset comprising over 400 metal clusters and 90 topologies~\cite{lee2021computational}. Fig.~\ref{fig:4} illustrates the distribution of frequently occurring building blocks and topologies among top-performing candidates, as determined by the proposed fitness function. Notably, the metal cluster N262 and topology \emph{bcg} were present in 33.8\% and 35.9\% of the top photocatalysts, respectively.

\begin{figure}[!t]
\centering
\includegraphics[width=\columnwidth]{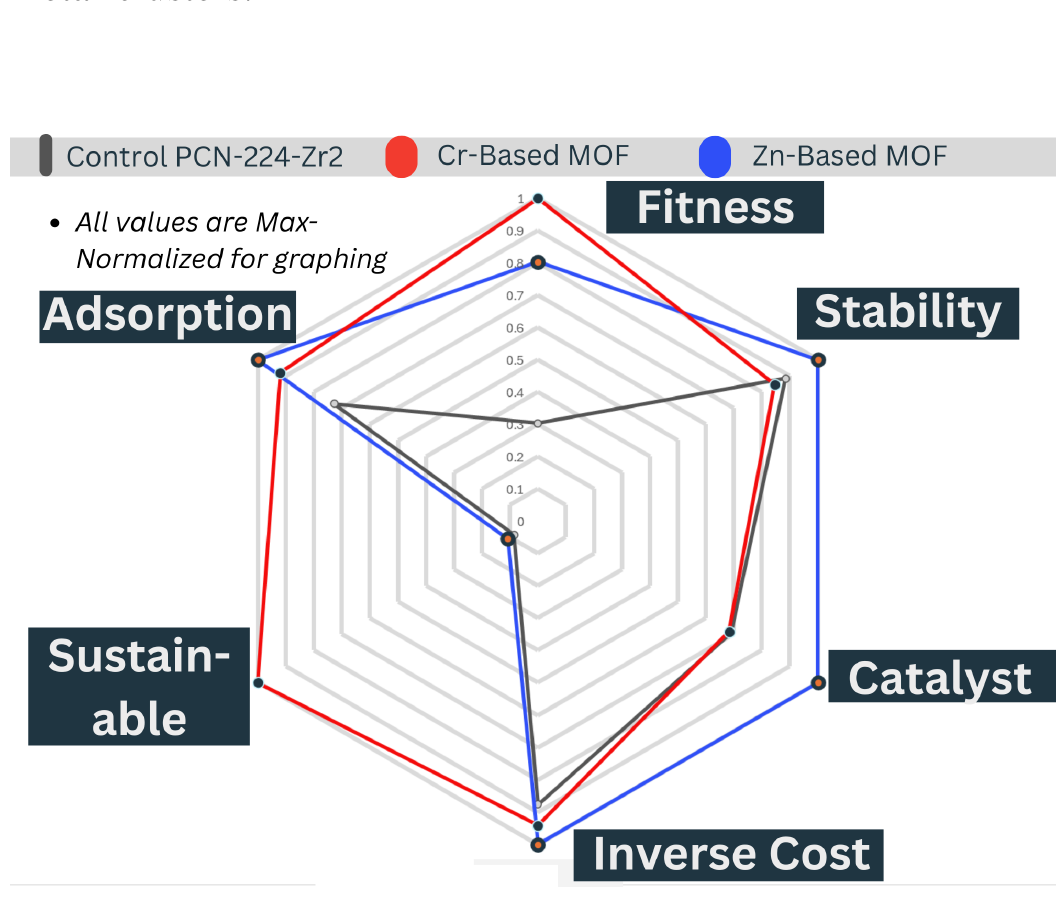}
\caption{Radar chart illustrating the performance comparison between the control MOF (PCN-224-Zr) and the two designed structures across five evaluation metrics. The Zn-based MOF achieved the highest score in four out of five aspects and exceeded the control in all measured attributes.}
\label{fig:5}
\end{figure}

Two particularly promising MOFs were designed using this approach: a chromium-based MOF and a zinc-based MOF. As shown in Fig.~\ref{fig:5}, the Cr-based MOF achieved the highest fitness score among all generated structures. Meanwhile, the Zn-based MOF demonstrated the highest fitness among those outperforming the most viable control material across all five key evaluation metrics. To assess the efficacy of the fitness function defined in Eq.~\ref{eq:fitness}, critical performance attributes associated with photocatalytic activity were quantitatively evaluated. These assessments confirmed a significant improvement in overall fitness and attribute-specific viability among selected MOFs.

\begin{figure}[!t]
\centering
\includegraphics[width=\columnwidth]{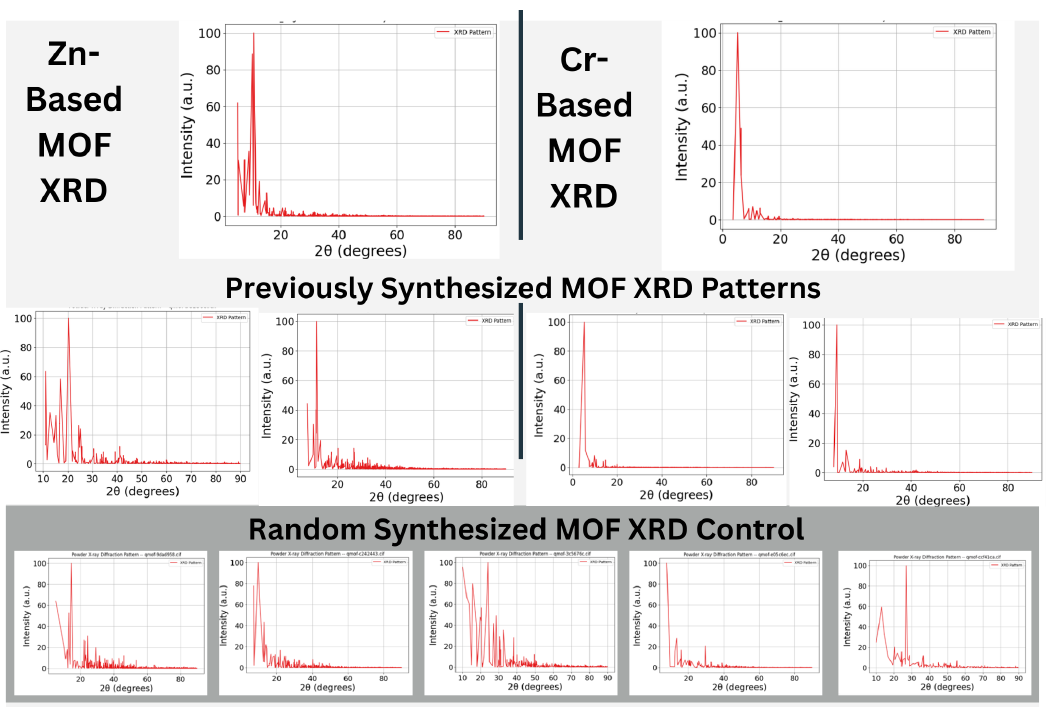}
\caption{X-ray diffraction (XRD) patterns of the two promising MOFs compared to those of previously synthesized structures from the QMOF database. For reference, XRD was also performed on five randomly generated MOFs, illustrating the broad variability in diffraction patterns across structures.}
\label{fig:6}
\end{figure}

X-ray diffraction (XRD) analysis was conducted for both the Cr- and Zn-based MOFs, along with a set of randomly constructed MOFs, as illustrated in Fig.~\ref{fig:6}. The random structures exhibited diverse diffraction profiles, with considerable variation in dominant peak position, peak count, area under the curve (AUC), and relative peak intensity. In comparison, both the Zn-based MOF and the Cr-based MOF show very similar pattern structures to previously synthesized materials, with almost identical dominant peak position, number of peaks, relative peak height, and comparable AUC.

\begin{figure}[!t]
\centering
\includegraphics[width=\columnwidth]{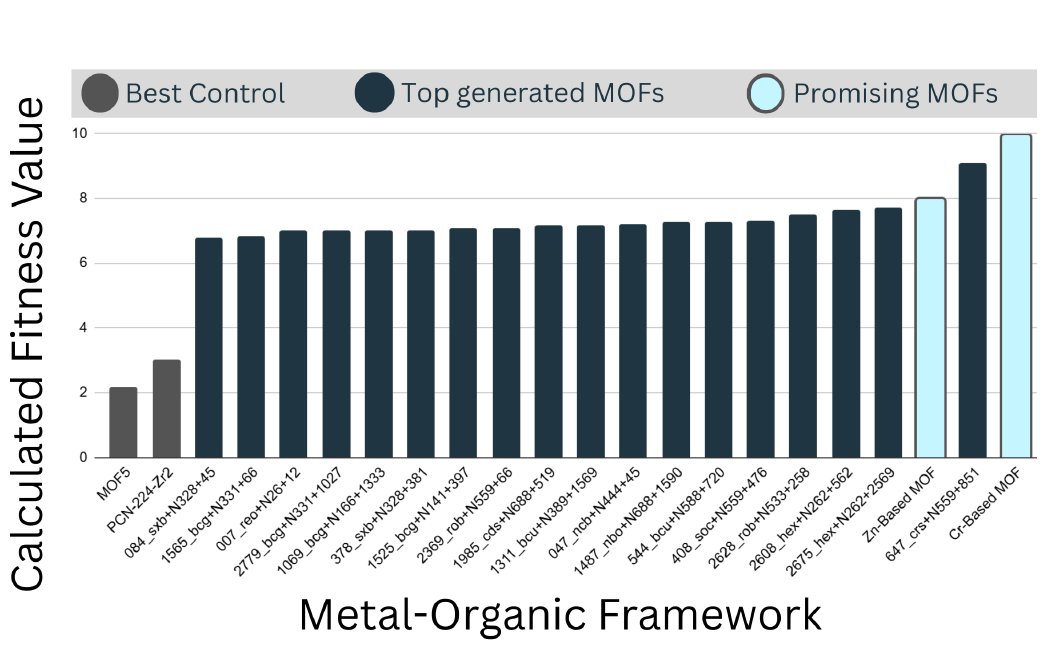}
\caption{Performance comparison between the 20 highest-ranked generated MOFs and the top-performing control structures. 20 candidates demonstrated over 125\% improvement in predicted photocatalytic viability.}
\label{fig:7}
\end{figure}

Furthermore, as presented in Fig.~\ref{fig:7}, 20 generated MOFs exhibit over 125\% improvement in predicted photocatalytic efficiency relative to the control material. Notably, the Zn-based MOF ranks among the top-performing candidates within this subset.

\section{Discussion}

\subsection{Result Interpretation}

These fitness values, significantly higher than MOFs found in existing literature, can be explained via two major reasons.

\begin{itemize}
\item \textbf{Limited prior optimization scope:} Previous studies are only able to optimize a narrow set of properties, prioritizing catalytic features due to their critical role in applications~\cite{yu2023computational,nazarian2015benchmarking}. Fig.~\ref{fig:5} illustrates, however, that the most significant gains were observed in traditionally overlooked properties such as sustainability, catalytic ability, and adsorption, with additional incremental improvements across other evaluated properties.
\item \textbf{Expansion of prior modelled aspects:} While catalytic ability has been a common focus in past research, this study extends the modeling approach by incorporating four distinct parameters to represent that attribute---substantially more than prior efforts~\cite{rosen2021machine}. The fitness function allows the model to effectively maximize these values, allowing for deeper analysis of common aspects.
\end{itemize}

Fig.~\ref{fig:4} explores the building blocks and topologies of the top 1,000 MOFs for photocatalysis. Distinct patterns found, such as the metal-cluster building block N262's consistent presence in the top compounds for photocatalysis, demonstrate that this work can identify underlying chemical traits and provide a means for future researchers to navigate complex chemical spaces. All of the presented metal clusters include a metal ion that is extractable from waste, such as electroplating sludge~\cite{boukayouht2023sustainable}. Alongside this, the top metal clusters also include some combination of hydrogen, carbon, nitrogen, or oxygen---four of the cheapest atoms in the designed dataset. Finally, the topologies all appear to be semi-tightly packed to ensure sound atomic packing for structural strength, but open, with significant organic linkers for higher available surface area to ensure extensive active sites for adsorption. Together, this demonstrates that this process can effectively uncover aspects deemed critical for application success, and narrows down the design space by highlighting those materials. This is especially influential in emerging fields (which inherently tend to be rarely modellable due to their novelty), where this process can elucidate viable materials for experimental work to focus on, while still demonstrating critical chemical concepts and patterns.

Fig.~\ref{fig:7} presents 20 metal-organic frameworks (MOFs) that outperform the best existing reference material for photocatalysis by over 125\%. These results demonstrate the potential of this approach to design materials that significantly exceed the performance of previously reported structures. All candidates meet various synthesis criteria, and the diversity in their structural components---reflected in their unique naming conventions---suggests strong practical potential for industrial implementation.

\subsection{Design Criterion Evaluation}

The integration of a Crystal Graph Convolutional Neural Network (CGCNN) model, a new fine-tuned CGCNN, combined with a proposed funnel-based filtering system, has led to significant gains in computational efficiency---achieving a speedup of approximately $6.54\times10^{7}$ times compared to traditional molecular modeling approaches~\cite{ajayi2024current}.

To address the challenge of feature overfitting during optimization, this work introduces two complementary techniques:
\begin{itemize}
\item \textbf{Funnel drop-off system:} At each of the five aspect levels, the bottom 5\% of candidates are eliminated based on performance. This strategy not only streamlines the dataset but also yields a 276\% improvement in computational efficiency by focusing resources on more promising materials.
\item \textbf{Multiplicative fitness function:} A multiplicative scoring system is used to evaluate material performance across five key features, since it strongly penalizes poor performance in any single feature, ensuring that only well-rounded candidates advance.
\end{itemize}

The proposed framework successfully identifies candidate materials that align with both performance and sustainability criteria. Notably, the Cr-based MOF contains approximately 12 times as many sustainably sourceable atoms in its structure compared to the most effective material in the control set. Both chromium and zinc have been documented as recyclable from industrial byproducts, such as electroplating sludge~\cite{boukayouht2023sustainable}, supporting environmentally responsible material sourcing.

As found in Fig.~\ref{fig:5}, among the generated candidates, the zinc-based material demonstrates superior overall performance across key metrics, including:
\begin{itemize}
\item \textbf{Structural stability:} 13\% improvement in stability metrics. This resilience ensures the preservation of active sites under high-pressure or humid environments, which is critical for maintaining consistent catalytic reactions throughout operational cycles. This reflects better mechanical properties and water resistance---common limitations in MOF materials.
\item \textbf{Catalytic performance:} 45\% increase in catalytic effectiveness, enabling robust photoelectric activity across the reaction cycle.
\item \textbf{Cost efficiency:} 14\% reduction in estimated cost relative to the control set, improving industrial applicability.
\item \textbf{Adsorption capabilities:} 37\% improvement in adsorption-related properties, such as increased reaction locations, improved selectivity for target molecules, and optimal removal of formed product.
\end{itemize}

To ensure real-world viability, environmental constraints---such as thermal stability above 148\textdegree~C (300\textdegree~F)---were also applied during material filtering through the funnel-based system~\cite{power2010101}. This further enhances the relevance of selected materials for deployment in demanding operational settings.

All proposed materials meet or exceed standard constructability thresholds in the field. Each structure contains fewer than 3,000 atoms, spans no more than 60~\AA\ in dimension, passes a Synthesis Accessibility Score $<6$, and maintains a Root Mean Squared Distance (RMSD) below 0.3~\cite{park2024inverse}. Additionally, all candidates passed the MOFDiff connectivity criterion~\cite{fu2023mofdiff}, were composed of components documented in previously synthesized structures~\cite{boukayouht2023sustainable}, and reached a natural, optimized state.

\subsection{Model Limitations}

Despite the effectiveness of the proposed framework, several limitations merit discussion.

First, the system relies heavily on machine learning models to predict material properties. While these models offer substantial improvements in computational efficiency~\cite{ajayi2024current}, their accuracy may not always align with physical reality~\cite{rosen2021machine}. This limitation can be mitigated through the incorporation of classical simulation techniques, such as Density Functional Theory (DFT)~\cite{nazarian2015benchmarking}, to verify the predicted results. In addition, promising candidates identified by the model can be validated through experimental synthesis to further ensure reliability. The integration of synthesizability constraints in the workflow also increases the likelihood of practical applicability.

Second, the model's performance is tied to a fitness function that combines multiple feature scores into a single optimization objective. An improperly weighted fitness function may unintentionally bias the selection of high-performing materials. However, results from this work suggest that the system is relatively robust to such imbalances. As demonstrated in Fig.~\ref{fig:5}, the zinc-based MOF outperforms control materials across all relevant metrics, indicating that even when certain attributes are misweighted, this procedure creates a material superior in every relevant feature. This would ultimately lead to a higher fitness regardless of the fitness function itself.

Several data-related constraints were encountered throughout this work; however, the framework consistently produced viable material designs despite these limitations. This outcome highlights the robustness of the proposed workflow in handling complex applications where data scarcity or variability may be present.

\begin{itemize}
\item \textbf{Small dataset limitations:} Certain prediction models, such as the water stability module, were trained on relatively small datasets (e.g., only 36 samples). Despite this, careful curation and training resulted in high predictive accuracy.
\item \textbf{Original data:} For several key features (e.g., cost and sustainability), data were derived from existing literature or manually curated experimental sources~\cite{boukayouht2023sustainable,satyak2024github}. While this expands the model's scope, it may introduce inconsistencies or assumptions that affect generalizability. Models such as the CGCNN have means of mitigating this, still capturing the underlying patterns and resulting in high predictive accuracy.
\item \textbf{Mixed optimization objectives:} Some features, such as band gap or Heat of Adsorption (HOA), do not follow a ``maximize is best'' rule~\cite{wu2017co2reduction}. The proposed framework addresses this by customizing evaluation logic based on whether a feature benefits from minimization or optimization within a specific range.
\end{itemize}

\subsection{Future Work}

Future research can explore the integration of a genetic algorithm~\cite{gustafson2019intelligent} applied to the remaining 10,986 MOFs. This would provide a substantially more diverse and robust base set compared to previous efforts, enabling the design of additional high-performing materials using the existing fitness evaluation framework.

Additionally, the development of a multi-output predictive model capable of estimating all 13 material features simultaneously presents a promising direction. Such an approach could lead to a significant improvement in computational efficiency---potentially up to 10 times faster than current methods. This would scale the workflow to support the rapid evaluation and design of tens of millions of candidate materials. To mitigate feature-specific overfitting, a funnel-based filtering system can be applied post-prediction to maintain balance and reliability across all target attributes.

\subsection{Relevance to the Field}

This research presents a versatile and scalable framework for computationally modeling metal-organic frameworks (MOFs) across a wide range of applications using a comprehensive fitness function. The proposed reinforcement-learning-based workflow not only facilitates the automated design and selection of high-performing structures but also allows for adaptability to specific application requirements. While demonstrated here for photocatalysis, the approach is readily extendable to other complex domains, such as drug delivery~\cite{gupta2019peg} and energy storage~\cite{chu2019metal}. Furthermore, the selection of specific features may be altered to fit the material usage situation, allowing for adaptability of this workflow across various implementations (e.g., pre-combustion compared to post-combustion photocatalysis) or environments.

This workflow is able to dissect patterns in the top MOFs for various applications, and provides a method for future work to investigate these promising building blocks. This has immediate influence in emerging technologies where the underlying mechanisms may not be well explored. The model's capability for multi-objective optimization distinguishes it from previous methods, effectively balancing trade-offs among diverse material properties. The funnel-based selection strategy further mitigates feature overfitting, contributing to the discovery of robust, high-performing MOFs. Overall, this framework enables deeper insights into material behavior and offers a data-driven pathway for accelerating innovation in MOF-based technologies.

\section{Conclusion}

This study introduces an efficient and scalable approach for the modeling and design of materials tailored for applications traditionally considered too complex for classical simulation techniques. Leveraging a comprehensive fitness function that incorporates over 260\% more features than prior work, the proposed framework advances multi-objective optimization in materials discovery.

This work presents an integrated workflow that combines a modified MOFReinforce model for feature-optimized MOF generation with Crystal Graph Convolutional Neural Networks (CGCNN) for accurate property prediction, linking generative and predictive models within a unified framework. A key contribution is the design and implementation of a funnel-based selection mechanism which further enhances computational efficiency, yielding a 276\% improvement compared to conventional high-throughput search strategies. The successful design of a zinc-based MOF---demonstrated to outperform the best control structure across multiple metrics---validates the robustness of this approach, even in the presence of potential fitness function inaccuracies. The model's ability to consistently identify key building blocks (e.g., the recurrent presence of linker N262 in high-performing structures) also indicates its capacity to uncover underlying chemical insights.

This work presents a generalizable and adaptable framework, with implications extending beyond photocatalysis to domains such as drug delivery and photothermal cancer therapies. The fitness function allows for flexible prioritization of feature sets---e.g., pre-combustion vs. post-combustion photocatalytic requirements---enhancing applicability in industrial contexts. All proposed materials meet synthesizability criteria, illustrating the high potential for future utilization. By addressing the challenge of ``unmodellable'' chemical systems, this framework opens pathways for the design of practical, scalable, and high-performing materials for a wide range of emerging and industrial applications.

\section*{Acknowledgment}

Special thanks to Professor Yamil Col\'on (University of Notre Dame) for his assistance as mentor for this project. Thanks to Professor Rosen (Princeton University), Professor Snyder (University of Illinois, Urbana-Champaign), Professor Vogiatzis (University of Tennessee, Knoxville), Professor Jaffe (University of Notre Dame), Dr.\ Meihaus (University of California, Berkeley), and Dr.\ DeMello (Stanford University Online High School) for their time in answering questions.

\end{document}